\documentclass[fleqn,12pt]{article}
\usepackage{amsmath}
\usepackage{tabls}
\usepackage{amsfonts,latexsym,stmaryrd}
\numberwithin{equation}{section}
\newcommand{\ii}{\mathrm{i}}

\newcommand{\pd}{\partial}
\newcommand{\dd}{\mathrm{d}}

\newcommand{\e}{\mathrm{e}}
\newcommand{\ket}[1]{\left|#1\right\rangle}
\newcommand{\bra}[1]{\left\langle #1\right|}

\newcommand{\I}{\mathbb{I}}

\newcommand{\ft}[2]{{\textstyle\frac{#1}{#2}}}
\def\tilde{\widetilde}
\def\1bar{1\hskip -.275cm -}
\def\2bar{2\hskip -.275cm -}
\def\3bar{3\hskip -.275cm -}

\newsavebox{\uuunit}

\newcommand{\nn}{\nonumber}

\newcommand{\nc}{\newcommand}
\nc{\la}{\lambda} \nc{\alf}{\alpha} \nc{\tht}{\theta}
\nc{\eps}{\epsilon} \nc{\ga}{\gamma} \nc{\Ga}{\Gamma}
\nc{\De}{\Delta} \nc{\de}{\delta} \nc{\si}{\sigma}
\nc{\ka}{\kappa} \nc{\om}{\omega} \nc{\qq}{\quad\quad}
\nc{\nf}{\infty} \nc{\dl}{\mathop{\smash{\cal L}}}
\nc{\ol}{\overline} \nc{\beq}{\begin{equation}}
\nc{\barr}{\begin{array}} \nc{\earr}{\end{array}}
\nc{\eeq}{\end{equation}} \nc{\beqa}{\begin{eqnarray}}
\nc{\dst}{\displaystyle}\nc{\pt}{\partial}
\nc{\eeqa}{\end{eqnarray}} \nc{\nnb}{\nonumber}
\nc{\bs}{\backslash}        \nc{\mbb}{\mathbb}
\nc{\brm}{\begin{remunerate}} \nc{\erm}{\end{remunerate}}
\nc{\vareps}{\varepsilon} \nc{\tb}{\tilde\beta_0} \nc{\ts}{\tilde
s} \nc{\tth}{\tilde \theta}
\newcounter{muni}
\usecounter{muni}
  \nc{\lapdec}{\mathop{\Delta}}

\def\theequation{\arabic{section}.\arabic{equation}}
\newenvironment{remunerate}{\begin{list}{{\rm \arabic{muni}.}}
{\usecounter{muni}
\setlength{\leftmargin}{0pt}\setlength{\itemindent}{38pt}}}{\end{list}}
\newcommand{\alg}[1]{\mathfrak{#1}}

\newcommand{\alG}{\alg{g}}

\newcommand{\alSU}{\alg{su}}

\newcommand{\alPSU}{\alg{psu}}

\nc{\cre}{\color[rgb]{1.00,0.00,0.00}}
\nc{\cgr}{\color[rgb]{0.00,1.00,0.00}}

\newcommand{\dg}{^\dagger}

\newcommand{\be}{\begin{equation}} \newcommand{\ee}{\end{equation}}
\newcommand{\bea}{\begin{eqnarray}} \newcommand{\eea}{\end{eqnarray}}
\newcommand{\ben}{\begin{displaymath}}
\newcommand{\een}{\end{displaymath}}

\makeatletter
\def\hlinewd#1{%
\noalign{\ifnum0=`}\fi\hrule \@height #1 %
\futurelet\reserved@a\@xhline} \makeatother

\begin{document}

\title{Sigma model from SU$(1,1|2)$ spin chain}

\author{ {\bf S. Bellucci, P.-Y. Casteill} \\
{\it INFN -- Laboratori Nazionali di Frascati,}\\
{\it Via E. Fermi 40, 00044 Frascati, Italy}
}

\maketitle
\begin{abstract}
We derive the coherent state representation of the integrable spin
chain Hamiltonian with supersymmetry group $SU(1,1|2)$. By the use
of a projected Hamiltonian onto bosonic states, we give explicitly
the action of the Hamiltonian on $SU(2)\times SL(2)$ coherent
states. Passing to the continuous limit, we find that the
corresponding bosonic sigma model is the sum of the known $SU(2)$
and $SL(2)$ ones, and thus it gives a string spinning fast on
$S^1_{\phi_1}\times S^1_{\varphi_1}\times S^1_{\varphi_2}$ in
$\rm{AdS}_5 \times S^5$. The full sigma model on the supercoset
$SU(1,1|2)/SU(1|1)^2$ is given.

\end{abstract}
\newpage

\tableofcontents
\section{Introduction}

The AdS/CFT
correspondence\cite{Maldacena:1998re,Gubser:1998bc,Witten:1998qj}
between strings on anti-de Sitter (AdS) spaces and boundary gauge
theories has generated much interest in recent years. One of the
most studied examples relates string theory on AdS$_5\times S^5$
to ${\cal N}=4$ supersymmetric Yang-Mills (SYM) gauge theory.
String states in the bulk are dual to gauge invariant operators in
the boundary and an increasing holographic dictionary between
their correlation functions has been derived. In
\cite{Berenstein:2002jq}, this holographic correspondence was
established in the neighborhood of null geodesics of AdS$_5\times
S^5$, where the geometry looks like a pp-wave \cite{Blau:2001ne}.
On such a geometry, string theory is known to be solvable
\cite{Metsaev:2001bj,Metsaev:2002re}, while on the gauge theory
side it corresponds to SYM operators with large ${\cal
R}$-symmetry charge $J$. In
\cite{Gubser:2002tv}\nocite{Frolov:2002av,
Tseytlin:2003ac,Frolov:2003qc,Frolov:2003tu,Beisert:2003xu,Frolov:2003xy,
Arutyunov:2003uj,Beisert:2003ea}-\cite{Frolov:2004bh}, the authors
studied the fluctuations around semiclassical spinning strings and
showed that, there also, energies of classical string solutions
matched anomalous dimensions of SYM operators with large charges.
Recently the matrix model approach to the anomalous dimension
matrix in ${\cal N}=4$ SYM theory was considered
\cite{Bellucci:2004rua}.

On the gauge theory side, the planar one-loop anomalous dimensions
in ${\cal N}=4$ SYM turned out to be described by integrable spin
chain Hamiltonians
\cite{Minahan:2002ve,Beisert:2003jj,Beisert:2003yb}. Thanks to
integrability, strong progress was then achieved in the comparison
of the spectrums on both sides, as it allowed to use powerful
Bethe Ansatz techniques\footnote{When non-planar
corrections are included
\cite{Beisert:2002ff}\nocite{Beisert:2003tq,Bellucci:2004ru,Bellucci:2004qx,Peeters1,
Bellucci:2005ma1,Peeters2}-\cite{Bellucci:2005ma2}, there arises
an interesting question, about the possibility of an extended or
modified notion of (quasi) integrability.}. We refer the
reader to \cite{tse1}\nocite{beis,tse2,zar,swa}-\cite{plefka} for
extensive reviews and citations and to
\cite{Beisert:2005di,Beisert:2005fw} for recent important results
on the subject where the authors give the Bethe Ans\"atze for the
full $SU(2,2|4)$ group in the thermodynamic limit at one and
higher loops.

 In the
continuous (BMN) limit, where $J\gg1$, the spin chains can be
identified with the worldsheet of closed strings\footnote{ For a
treatment of the problem of fermion doubling and the BMN
correspondence, see
\cite{Bellucci:2004rub,Bellucci:2004ruc,Bellucci:2004rud}.}. The
spin chain excitations give then the string profile in the
symmetry group taken as target space and the spin chain
Hamiltonian describes the dynamic of the string. As for the BMN
case, the perturbative regime of SYM is accessible to this limit
and accordingly, string and spin chain sigma model actions should
agree. By the use of a coherent vector description of the spin
chain, this was shown to be the case in \cite{Kruczenski:2003gt}
for the $SU(2)$ subsector of the theory. Extension to the whole
$SO(6)$ and its other compact subgroups was then performed in
\cite{Kruczenski:2004kw}\nocite{Hernandez:2004uw,Stefanski:2004cw,
Kristjansen:2004za}-\cite{Kruczenski:2004cn}. The non compact
$SL(2)$ case was studied in
\cite{Stefanski:2004cw,Bellucci:2004qr,Ryang:2004pu}. This study
was extended to the supersymmetric sectors $SU(1|2)$
\cite{Hernandez:2004kr} , $SU(1,1|1)$ \cite{Bellucci:2005vq} and
$SU(2|3)$ \cite{Hernandez:2004kr,Stefanski:2005tr}. In this last
paper, the authors also discussed on a generalisation to the full
$SU(2,2|4)$.

In all cases, semiclassical spinning string states were identified
with coherent states. These are built from spin chain states by
acting with the coset $G/H_{\ket{0}}$ on a vacuum $\ket{0}$, with
$G$ the subsector studied and $H_{\ket{0}}$ the stabilizer of $G$
by respect to $\ket{0}$. Because of the properties of the coherent
states, one can go to a path integral formulation without loosing
any information of the initial theory. Passing then to the
continuous limit along the spin chain gives the sigma model. We
refer the reader to
\cite{Kazakov:2004nh}\nocite{Kruczenski:2004wg,Arutyunov:2004yx,Park:2005ji,Khan:2005fc,
Berenstein:2005fa,Freyhult:2005fn,
Beisert:2005mq,Hernandez:2005nf,Ryang:2005yd,Schafer-Nameki:2005tn,Schafer-Nameki:2005is,
koch}-\cite{Minahan:2005mx} for further developments in this
subject.

We will focus in this paper on the $SU(1,1|2)$ sector of the
theory. This sector is interesting as it generalizes the two
simpler bosonic sectors $SU(2)$ and $SL(2)$. Each of these two
subsectors carry information from the two main bosonic parts of
$SU(2,2|4)$ which are $SO(2,4)$ and $SO(6)$, and interact between
themselves via supersymmetric charges. The $SU(1,1|2)$ sector
corresponds to SYM operators made out of two scalars carrying
different $SU(2)$ charges and two fermions, plus derivatives along
a fixed direction ($SL(2)$ charge). The corresponding Bethe Ansatz
in the thermodynamic limit is discussed
in details in \cite{Beisert:2005di,Beisert:2005fw}.
The sector is non compact,
and its representations are thus infinite dimensional. We first
derive a coherent state representation of the spin chain
Hamiltonian. Like in the simpler $SU(1,1|1)$ case
\cite{Bellucci:2005vq}, the Hamiltonian results to
have a non-linear form, with a logarithmic term.
But moreover, it cannot be expressed, as it was the case for the
$SU(1,1|2)$ subgroups, in terms of the square on the superspace of a single vector
$(\vec{n}_2-\vec{n}_1)^2$ built from coherent states. This makes its
fermionic part quite involved. However, by passing to the
continuous limit where expressions simplify a lot, the fermions
reassemble in the simple square $\pd \vec{n}^2$, like in
\cite{Hernandez:2004kr,Bellucci:2005vq,Stefanski:2005tr}. The so obtained
sigma model should then correspond to a string moving on the
supercoset $SU(1,1|2)/SU(1|1)^2$. We check that this is at least the case
for the bosonic action that turn out to be the sum of the $SU(2)$ and $SL(2)$
sigma models \cite{Kruczenski:2003gt,Stefanski:2004cw,Bellucci:2004qr}.

The paper is organized as
follows. In Section \ref{scoh} we build the SU$(1,1|2)$ coherent
state. In Section \ref{shamact}, we give an expression of
the two-site Hamiltonian suitable for acting on our coherent states.
In Section \ref{sham} we derive the sigma model on the group manifold $SU(1,1|2)/SU(1|1)^2$.
As the fermionic part of the Hamiltonian action on coherent states
resulted into a too long and too complicated expression to be presented in a paper (69 terms),
we do this in two steps. First, we focus in Subsection \ref{shamtrunc}
on a truncated Hamiltonian acting on $SU(2)\times SL(2)$ spin chains. It consists
in the projection of the full $SU(1,1|2)$ Hamiltonian onto pure bosonic states.
Its action on two coherent states is given explicitly and gives rise to
a strange mix between the $SU(2)$ and $SL(2)$ separate actions.
Then in Subsection \ref{ssigmamodel}, we go to the continuous limit,
putting back the fermions.
In Section \ref{sstring} we show that the bosonic sigma
model arises when considering a string spinning fast on
$S^1_{\phi}$ on AdS$_5$ and $S^1_{\varphi_1}\times S^1_{\varphi_2}$ in $S^5$. Finally in
Section \ref{scon} we summarize our results. Appendix
\ref{aa} collect the definitions in terms of oscillators of states and charges,
as well as the commutation relations of the $SU(1,1|2)$ algebra.

\section{The coherent state}
\label{scoh}

Coherent states are defined by the choice of a group $G$ and a
vacuum  $\ket{0}$ in a representation ${\cal R}$ of the group. We
denote by $H_{\ket{0}}$ the corresponding stabilizer subgroup, \emph{i.e.}
the group of elements of $G$ that leave $\ket{0}$ invariant up to
a phase. The coherent state is then defined by the action of a
finite group element of $g\in G/H_{\ket{0}}$  on $\ket{0}$. With $G$ chosen
as being $SU(1,1|2)$, we will take $\ket{0}$ to be the physical
vacuum $\ket{\phi_0}$. The generators for the algebra $\alG$ are
taken to be
$$T_A=(J_0,R_0,P,K,R_{23},R_{32},Q_2,Q_3,S_2,S_3,\dot{Q}_2,\dot{Q}_3,\dot{S}_2,\dot{S}_3
)~. $$
Conventions and details about the algebra and its singleton
representation are given in appendix A. The stabilizer subgroup
$H_{\ket{\phi_0}}$ is generated by
$$H_{\ket{\phi_0}}=\{J_0,R_0,Q_3,S_3,\dot{Q}_2,\dot{S}_2 \}=SU(1|1)^2~~~.$$
We chose the coherent state to be \bea\label{coh-st:gen}
\ket{\vec{n}}&=&g(\vec{n})\ket{\phi_0}=\e^{z R_{3,2}- \bar{z}
R_{2,3}}\e^{u P- \bar{u} K}\, \e^{-\xi Q_2-\bar{\xi} S_2}\,
\e^{-\theta \dot{Q}_3-\bar{\theta} \dot{S}_3}\, \ket{\phi_0}~~,
\eea where $z=\psi\, \e^{\ii\,\varphi}$ and $u=\rho\,
\e^{\ii\,\phi}$. It is parameterized by four real parameters
$\rho$, $\psi$, $\phi$, $\varphi$ and two complex Grassmann
variables $\xi$ and $\theta$. The coherent state $\ket{{\bf n}}$
of the full spin chain writes as the direct product \be
 \ket{{\bf n}}=\ket{\vec{n}_1}\otimes \ket{\vec{n}_2}\ldots \ket{\vec{n}_J}
\label{kcoh}
\ee
where each $\ket{\vec{n}_k}$ denotes a coherent state
(\ref{coh-st:gen}) with its own parameters $\rho_{k}(t),~\psi_{k}(t),$
...~, and describes the spin chain excitation at site $k$ and time
$t$.

The spin chain action will then be given in terms of the spin chain Hamiltonian $H$
by
\bea
S &=& -\int \dd t \left( \ii\,\bra{\bf n} \partial_t \ket{\bf n}+
{\hat \lambda} \bra{\bf n} H \ket{\bf n}   \right),\label{action}
\eea
which, after taking the continuous limit $J\rightarrow\infty$, will lead us to the
sigma model.

~\newline
The first task consists in expending the coherent states in
 the basis $\{\ket{\phi_m},\ket{Z_m},\ket{\lambda_m},\ket{\mu_m}\}$, with
 $\phi$,  $Z$ being scalar fields, $\lambda$, $\mu$ being fermions, and
 $m=0,1,2,\ldots $ labelling the number of derivatives among a fixed direction.
This is done using Table \ref{table:actstates} and gives the following expression
for the coherent state expression~:
\bea
  \ket{\vec{n}}&=&\sum_{m=0}^{\infty}\frac{\e^{\ii\,\,m\,\phi}\,
\tanh^m\!\rho}{\cosh\!\rho}
  \bigg[ (1+\ft12\,\xi\bar{\xi}+\ft12\,\theta\bar{\theta}+\ft14\,\xi\bar{\xi}\theta\bar{\theta})
\,(\cos\psi\ket{\phi_m}+\e^{\ii\,\varphi}\sin\psi\ket{Z_m})\nn\\
&& \phantom{\sum_{m=0}^{\infty}\frac{\e^{\ii\,\,m\,\phi}\,
\tanh^m\!\rho}{\cosh\!\rho}}+{\e^{-\ii\,\phi}\xi\,\theta\over \cosh\!\rho}
\left({m\over\sinh\!\rho}-
\sinh\!\rho\right)\,(\cos\psi\ket{Z_m}-\e^{-\ii\,\varphi}\sin\psi\ket{\phi_m})\nn\\
&&\phantom{\sum_{m=0}^{\infty}\frac{\e^{\ii\,\,m\,\phi}\,
\tanh^m\!\rho}{\cosh\!\rho}}-{\xi\over \cosh\!\rho}(1+\ft12\,\theta\bar{\theta})
\,\ket{\mu_m}-{\theta\over \cosh\!\rho}(1+\ft12\,\xi\bar{\xi})\,\ket{\lambda_m}\bigg]~~.\label{cm}
\eea
The coherent states $\ket{\vec{n}}$, although not orthogonals, are normalized~:
$$\bra{\vec{n}\, }\vec{n}\rangle=1~.$$
They are over-complete, \emph{i.e.} they fulfill a resolution of unity :
\begin{equation}\label{res-un}
  \I=\frac{2j_{sl2}-1}{4\,\pi^2}\int_0^{2\pi}\dd\phi
  \int_0^{2\pi}\dd\varphi
  \int_0^{\ft\pi2}\sin\!2\psi\,\dd\psi
  \int_0^\infty\sinh\!2\rho\,\dd\rho
  \int\dd\xi\dd\bar{\xi}\dd\theta\dd\bar{\theta}\,
  \ket{\vec{n}}\bra{\vec{n}}
~.
\end{equation}
Here, $j_{sl2}$ is the spin by respect to the $sl(2)$ subalgebra.
In our choice of states, it equals $\ft12$ for the bosons and 1
for the fermions. In an arbitrary $j_{sl2}$ representation, the
coherent state infinite expansion depends explicitly on $j_{sl2}$
\cite{book:perelomov}, and the integral over $\dd\rho$ gives a
$(2j_{sl2}-1)^{-1}$ factor. Therefore, when acting on bosons, the
integral should be seen as acting on coherent states
$\ket{\vec{n}(j_{sl2})}$ \emph{in the limit}
$j_{sl2}\rightarrow\ft12$ (see \cite{Bellucci:2004qr} for
details).

~\newline
It is possible to associate to each coherent state a
point $\vec{n}=\{n_{J0},n_{R0},\cdots\}$ in the superspace by
defining
\begin{equation}
n_A \equiv\bra{\vec{n} }\, T_A \, \ket{\vec{n}}~.\label{na}
\end{equation}
The action of the charges is given in Table \ref{table:actstates} and
leads, after summation in (\ref{cm}) to
\bea\vec{n}~:~
\left\{\begin{array}{rl}
n_{J_0}=&\ft12\,\cosh\!2\rho\,(1- \xi\bar{\xi}- \theta\bar{\theta})  \\[3mm]
n_{R_0}=&\ft12\,\cos 2\psi\,(1+\xi\bar{\xi}+\theta\bar{\theta})\\[3mm]
n_{P}=& \overline{n_K}=\e^{-\ii \, \phi } \, \cosh\rho\, \sinh\rho \left ( 1-\xi \bar{\xi }-\theta \bar{\theta }  \right )  \\[3mm]
n_{R_{23}}=& \overline{n_{R_{32}}}=~\e^{\ii \, \varphi } \, \cos\psi\, \sin\psi \left ( 1+\xi \bar{\xi }+\theta \bar{\theta } \right )  \\[3mm]
n_{Q_2}=& \overline{n_{S_2}}=~\cos\psi\, \cosh\rho\ \bar{\xi }
-\e^{-\ii \,(\phi +\varphi)} \,\sin\psi\, \sinh\rho\ \theta   \\[3mm]
n_{Q_3}=&  \overline{n_{S_3}}=~\e^{\ii \, \varphi }\, \sin\psi \, \cosh\rho\ \bar{\xi }
+\e^{-\ii \, \phi } \, \cos\psi \, \sinh\rho\ \theta \\[3mm]
n_{\dot{Q}_2}=& \overline{n_{\dot{S}_2}}= \e^{-\ii \, \phi } \, \cos\psi\, \sinh\rho\ \xi
 -\e^{\ii \, \varphi } \, \sin\psi\, \cosh\rho\ \bar{\theta } \\[3mm]
n_{\dot{Q}_3}=& \overline{n_{\dot{S}_3}}= \e^{-\ii \, ( \phi +\varphi) }  \, \sin\psi\, \sinh\rho\ \xi
+\cos\psi\, \cosh\rho\,\bar{\theta }
\end{array}\right.~~.\label{ncomp}
\eea
  The resulting vector is null $n^A n_A=0$ with respect to the
 metric $g_{AB}$ given by
\bea
n_A m^A&=&g^{AB} n_A m_B\label{metab}\\
&=& \ft12\,n_{J_0}\,m_{J_0}- \ft12\,n_{R_0}\,m_{R_0}
-\ft12\,n_{P}\,m_{K}-\ft12\,n_{R_{23}}\,m_{R_{32}}-\ft12\,n_{Q_i}\,m_{S_i}-\ft12\,\,n_{\bar{Q}_i}\,m_{\bar{S}_i}
+{\rm h.c.}\nn \eea Contrary to the usual case, the metric is not
defined through the Killing metric, which here vanishes
identically, but is given by the Casimir of the group (see
Appendix A).

~\newline
The first (Wess-Zumino) term in (\ref{action}) can be easily evaluated
by taking the derivative of (\ref{cm}) and then performing the infinite sum.
It has the simple form
\bea
\ii \bra{\bf n} \partial_t\! \ket{\bf n} &=&
\sum_{\rm{sites~}k} \left[-\sinh^2\!{\rho}\ \partial_t \phi-\sin^2{\psi}\ \partial_t \varphi
+\ft{\ii}{2}(\bar{\xi}\,D_t\xi+\xi\,\bar{D}_t\bar{\xi}
+\bar{\theta}\,D_t\theta+\theta\,\bar{D}_t\bar{\theta})
\right]_k\nn\\
\label{ndn}
\eea
with the covariant derivative defined as
\be
D_a \equiv \pd_a+\ii\, C_a~,
\quad\quad C_a\equiv \sinh^2\!{\rho}\ \partial_a \phi-\sin^2{\psi}\ \partial_a \varphi~.
\label{WSZ0}
\ee
Evaluating the second term in (\ref{action}) requires much more work. In order
to compute the average of the Hamiltonian between two spin chain coherent states $\ket{\bf n}$,
one should first express the Hamiltonian action on the basis
$\{\phi_m,Z_m,\lambda_m,\mu_m\}$. This is done in the next section.

\section{Hamiltonian action on $SU(1,1|2)$ states}
\label{shamact}

We here rewrite the SU$(1,1|2)$ two-sites harmonic Hamiltonian given by Beisert in
\cite{Beisert:2003jj} in the $\{\phi_m,Z_m,\lambda_m,\mu_m\}$ basis. The total Hamiltonian on the
spin chain is then given by the summation over all two neighboring sites along the spin chain~:
$$
  H=\sum_{k=1}^{J}H_{k\,k+1}~~~.
$$
Omitting the site's $k$ indices, a two-sites state is given by
$ \ket{A_m,B_n}\equiv\ket{A_m}\otimes\ket{B_n}$,
where $A_m$, $B_m$ stand for any of the $\{\phi_m,~Z_m,~\lambda_m,~\mu_m\}$. We have also to introduce
some other definitions, accounting for the usual harmonic number $h(m)$, a permutation operator\footnote{
Let us point out here that $\cal P$ permutes only letters, not their $sl(2)$ charges, and
it does \emph{not} take into account supersymmetric gradings, as it is common in the literacy.}
$\cal P$, raising/lowering operators ${\cal T}_{1,2}^\pm$, and a supersymmetric operator $\cal Q$~:
$$
\begin{array}{l}
h(m)=\sum_{i=1}^m\frac1i~~,\quad\quad\quad {\cal P}\ket{A_m,B_n}=\ket{B_m,A_n}~~,\\[3mm]
{\cal T}_{1}^\pm\ket{A_m,B_n}=\ket{A_{m\pm1},B_n}~~,~~{\cal T}_{2}^\pm\ket{A_m,B_n}=\ket{A_m,B_{n\pm1}}~~,\\[3mm]
{\cal Q}\ket{A_m,B_n}=\ket{{\cal Q}(A)_m,{\cal Q}(B)_n}~~,~~
{\cal Q}(\phi)=\lambda~~,~~{\cal Q}(\lambda)=\phi,~~{\cal Q}(Z)=-\mu,
~~{\cal Q}(\mu)=-Z~~.
\end{array}$$
By looking to the general shapes arising from the action of the harmonic Hamiltonian
\cite{Beisert:2003jj} on states
with a small number of derivatives, it is possible to
get a whole picture of its general action in the $SU(1,1|2)$ subsector. One ends with three possible cases :

\vspace*{5mm}$\bullet$ Boson/boson interaction
\bea
H_{12}\ket{A_k,B_l} =&&
\dst   \left[h(k)+h(l)+\frac{1-\cal P}{k+l+1}\right] \, \ket{A_{k},B_{l}}\nn\\
&+&\dst \sum_{i=1}^k \left[\frac{1-\cal P}{k+l+1}\left( 1 -\frac{{\cal Q} \,{\cal T}_{2}^-}{l+i}\right)
-\frac{1}{i}\, \right]\,\ket{A_{k-i},B_{l+i}}\label{Ham:pairBB}\\[3mm]
&+&\dst \sum_{j=1}^l \left[\frac{1-\cal P}{k+l+1}\left( 1 +\frac{{\cal Q} \,{\cal T}_{1}^-}{k+j}\right)
-\frac{1}{j}\, \right]\,\ket{A_{k+j},B_{l-j}}\nn
\eea
where letters $A$ and $B$ stand either for $\phi$ or $Z$.

\vspace*{5mm}$\bullet$ Fermion/fermion interaction
\bea
H_{12}\ket{A_k,B_l}&=&
\dst   \left[h(k+1)+h(l+1)-\frac{1-\cal P}{k+l+2}\left(1
+\left ( k+1\right )  \, {\cal Q} \, {\cal T}_{1}^{+}
-\left ( l+1\right )  \, {\cal Q} \, {\cal T}_{2}^{+}\right)\right] \, \ket{A_{k},B_{l}}\nn\\
&+&\dst \sum_{i=1}^k \left[\frac{\left(l+1\right)\left(1-{\cal P}\right)}{k+l+2}\left({\cal Q} \,{\cal T}_{2}^+ -\frac{1}{l+i+1}\right)
+\frac{1}{l+i+1}-\frac{1}{i}\, \right]\,\ket{A_{k-i},B_{l+i}}\label{Ham:pairFF}\\[3mm]
&+&\dst \sum_{j=1}^l \left[\frac{\left(k+1\right)\left(1-{\cal P}\right)}{k+l+2}\left(-{\cal Q} \,{\cal T}_{1}^+ -\frac{1}{k+j+1}\right)
+\frac{1}{k+j+1}-\frac{1}{j}\, \right]\,\ket{A_{k+j},B_{l-j}}\nn
\eea
where letters $A$ and $B$ stand either for $\lambda$ or $\mu$.

\vspace*{5mm}$\bullet$ Boson/fermion interaction
\bea
H_{12}\ket{A_k,B_l} =&&
\dst   \left[h(k+F_A)+h(l+F_B)-\left ( \frac{F_{B}}{1+k}+\frac{F_{A}}{1+l}\right )\,{\cal P}\right] \, \ket{A_{k},B_{l}}\nn\\
&+&\dst \sum_{i=1}^k \left(\frac{F_{B}-F_{A}\,{\cal P}}{l+i+1}
-\frac{1}{i}\, \right)\,\ket{A_{k-i},B_{l+i}}\label{Ham:pairBF}\\[3mm]
&+&\dst \sum_{j=1}^l \left(\frac{F_{A}-F_{B}\,{\cal P}}{k+j+1}
-\frac{1}{j}\, \right)\,\ket{A_{k+j},B_{l-j}}\nn
\eea
where letters $A$ and $B$ stand for any letter, $F_A$ and $F_B$ being their respective
supersymmetric grading ($F_\phi=F_Z\equiv0$ and $F_\lambda=F_\mu\equiv1$).
The condition $|F_A-F_B|=1$ is assumed.

~\newline

One can remark here that in going from the oscillator picture to precise states
$\{\phi_m$, $Z_m$, $\lambda_m$, $\mu_m\}$, the harmonic Hamiltonian \cite{Beisert:2003jj}
loses its very concise and elegant form. However, it gains two nice advantages~: first, the conditions
in the number of oscillators become implicit ; second, although more complicated, its
\emph{computation} can be done much faster. Indeed, the two-site harmonic Hamiltonian, because of its
sum on possible permutations on oscillator sites, is computable in exponential time. Its
derivation in the form given here is computable in linear time
by respect to the number of oscillator composing the initial states. Therefore, while
computations of states as \emph{e.g.} $H_{12}\ket{\phi_{10},\mu_{10}}$ were reaching
the capacities of normal computers, they become here immediate. Another expression
computable in quadratic time was given in \cite{Zwiebel:2005er} as the anti-commutator
of lowering/increasing length operators.

~\newline

\section{$SU(1,1|2)$ sigma model}
\label{sham}

The next task consists in taking the average of the Hamiltonian by two-sites coherent states :
\begin{equation}\label{average}
\bra{\vec{n}_1(\rho_1,\psi_1,\cdots),
\vec{n}_2(\rho_2,\psi_2,\cdots)}H_{12}
\ket{\vec{n}_1(\rho_1,\psi_1,\cdots),\vec{n}_2(\rho_2,\psi_2,\cdots)}~.
\end{equation}
The computation is extremely long and tedious, as one has to act with equations
(\ref{Ham:pairBB}), (\ref{Ham:pairFF}), (\ref{Ham:pairBF}) on two coherent states (\ref{cm}),
and then perform a double or triple sum. It is however still doable with the extensive use
of a computer.

In the $SU(1,1|1)$ case, it was possible to express the Grassmann
variables appearing in (\ref{average}) in a very compact form as
they just summed up with the bosonic ones into a logarithm. We
could not get a simple expression in the considered $SU(1,1|2)$
case. It seems that this is a direct consequence of the
supersymmetric mixing between $su(2)$ and $sl(2)$ subalgebras. For
the sake of simplicity, we will therefore just give the bosonic
part of the two-site Hamiltonian average. We will return to
fermionic considerations when taking the continuous limit, where
expressions simplifies a lot.

\subsection{$SU(2)\times SL(2)$ truncated Hamiltonian}
\label{shamtrunc}

In the $SU(2)$ subsector, the average (\ref{average}) was linear in $\left(\vec{n}_1
-\vec{n}_2\right)^2$ \cite{Kruczenski:2003gt}, while this same square
appeared\footnote{
It is always assumed that square acting on vectors use the corresponding group metrics $g_{AB}$.}
 in a logarithm in the $SL(2)$ \cite{Stefanski:2004cw,Bellucci:2004qr} and
$SU(1,1|1)$ \cite{Bellucci:2005vq} subsectors. As we will see in this subsection, the average
Hamiltonian in the $SU(1,1|2)$ sector cannot be expressed as a function of
$\left(\vec{n}_1-\vec{n}_2\right)^2$ only.

As all fermions appear in the coherent state (\ref{cm}) together
with a Grassmann variable, it is clear that the bosonic part of
the average Hamiltonian (\ref{average}) will be given by
(\emph{i}) restricting ourselves to coherent states with Grassmann
variables set to zero and (\emph{ii}) taking as Hamiltonian the
projection of the bosonic/bosonic interaction (\ref{Ham:pairBB})
onto bosonic states. Computation of (\emph{ii}) is
straightforward, and leads to \bea
H_{12}^{\rm{bosonic}}\ket{A_k,B_l} &=&
\dst   \big[h(k)+h(l)\big] \, \ket{A_{k},B_{l}}-\sum_{\tiny\begin{array}{l} i=0\\
i\neq k\end{array}}
^{k+l}\frac{1}{|k-i|}\,\ket{A_{i},B_{k+l-i}}\nn\\
&&+\dst \frac{1-\cal P}{k+l+1}\sum_{i=0}^{k+l}\,\ket{A_{i},B_{k+l-i}}\nn\\[3mm]
&=&
\dst H_{12}^{sl(2)}\ket{A_k,B_l} + \frac{H_{12}^{su(2)}}{k+l+1}\sum_{i=0}^{k+l}\,
\ket{A_{i},B_{k+l-i}}~~.\label{bosHam}
\eea
Here, $A$ and $B$ stand for $\phi$ or $Z$. $H_{12}^{sl(2)}$ is the $sl(2)$ two-sites
Hamiltonian found in \cite{Beisert:2003jj}, while $H_{12}^{su(2)}\equiv {1-\cal P}$ is the
usual Heisenberg XXX$_{1/2}$ two-sites Hamiltonian. The Hamiltonian (\ref{bosHam}) appears as
an Hamiltonian on $SU(2)\times SL(2)$ spin chains. By construction it commutes with all the
bosonic charges, and the spin $j$ states $\ket{j}$ defined in appendix A are still
eigenvectors~: $H_{12}^{\rm{bosonic}}\ket{j}=2 h(j)\ket{j}$ \cite{Beisert:2003jj}.

~\newline
Computing the average of $H_{12}^{\rm{bosonic}}$ in the bosonic sector of $SU(1,1|2)$ then leads to

\begin{equation}\label{averagebos}
\bra{\vec{n}_1,\vec{n}_2}H_{12}^{\rm{bosonic}}
\ket{\vec{n}_1,\vec{n}_2}=\left(1-\frac {(\vec{n}_2-\vec{n}_1)_{su2}^2}
{(\vec{n}_2-\vec{n}_1)_{sl2}^2}\right)\,\log(1-(\vec{n}_2-\vec{n}_1)_{sl2}^2) ~~~,
\end{equation}

~\newline
\noindent where $(\vec{n}_2-\vec{n}_1)_{sl2}^2$ and $(\vec{n}_2-\vec{n}_1)_{su2}^2$ are exactly the terms appearing in the $SU(2)$ and $SL(2)$ subsectors
respectively ! To be more precise, one has
$$(\vec{n}_2-\vec{n}_1)_{sl2}^2=
(\vec{n}_2-\vec{n}_1)^2\big|_{\psi,\varphi,\theta,\xi\rightarrow0}~~{\rm and}~~
(\vec{n}_2-\vec{n}_1)_{su2}^2=-(\vec{n}_2-\vec{n}_1)^2
\big|_{\rho,\phi,\theta,\xi\rightarrow0}
$$
so that
\bea
(\vec{n}_2-\vec{n}_1)_{sl2}^2&=&\frac{1}{2}\,\left(1-\cosh2 \, \rho_{1} \,
\cosh 2\rho_{2} +\cos\left(\phi_{1}-\phi_{2}\right)\,
\sinh2\,\rho_{1}\, \sinh2\,\rho _{2}\right )~~,\quad\quad\nn\\
(\vec{n}_2-\vec{n}_1)_{su2}^2&=&\frac{1}{2}\,\left(1-\cos2 \, \psi_{1}\,\cos2 \, \psi_{2}
 -\cos\left ( \varphi_{1}-\varphi_{2}\right )\,
 \sin2 \, \psi_{1}\, \sin2 \, \psi_{2} \right )~~~.
\eea
The limit to the subsectors $SU(2)$ and  $SL(2)$ appears clearly, as
they impose respectively $(\vec{n}_2-\vec{n}_1)_{sl2}^2\rightarrow0$ and
$(\vec{n}_2-\vec{n}_1)_{su2}^2\rightarrow0$.

Computing the square of the full coherent vectors
with $g_{AB}$, one finds
$$
(\vec{n}_2-\vec{n}_1)^2=(\vec{n}_2-\vec{n}_1)_{sl2}^2
-(\vec{n}_2-\vec{n}_1)_{su2}^2+~~{\rm fermions}~~~.
$$
Therefore, just looking to (\ref{averagebos}), one concludes that it will not be possible to
express the average of the full $SU(1,1|2)$ Hamiltonian just in terms of
$(\vec{n}_2-\vec{n}_1)^2$, as it was the case in the $SU(2)$ and
$SU(1,1|1)$ subsectors : the square on ``coherent
vectors'' is cut into two pieces, in order to give a mix of precedent $SU(2)$ and $SL(2)$
found shapes.

\subsection{Continuous limit}
\label{ssigmamodel}

The fermionic part of (\ref{average}) is much more involved. It contains
69 different terms without any clear structure. Even the quadratic
terms appear with coefficients that mix trigonometric and hyperbolic functions in an
highly non trivial way. We found no simplification neither rewriting
(\ref{averagebos}) with the Ansatz
$$\bra{\vec{n}_1,\vec{n}_2}H_{12}
\ket{\vec{n}_1,\vec{n}_2}=\left(1-\frac {(\vec{n}_2-\vec{n}_1)_{su2}^2+{\rm fermions}_1}
{(\vec{n}_2-\vec{n}_1)_{sl2}^2+{\rm fermions}_2}\right)\,
\log(1-(\vec{n}_2-\vec{n}_1)_{sl2}^2-{\rm fermions}_1)~~~,
$$
that was suggested by the bosonic average.

~\newline
Fortunately, everything simplifies a lot in the limit $\vec{n}_2\rightarrow\vec{n}_1$.
The result is, with fermionic part included,

$$\bra{\vec{n}_1,\vec{n}_2}H_{12}
\ket{\vec{n}_1,\vec{n}_2}=-\epsilon^2\ g^{AB}\ {\delta n}_A\ {\delta n}_B+{\cal O}(\epsilon^3)$$

~\newline
\noindent where $\vec{n}_2=\vec{n}_1+\epsilon\,\vec{\delta n}$. Like all results previously
found for $SU(1,1|2)$ subsectors, it appears that in the continuous limit, the Hamiltonian is
just the square, made with the corresponding superspace metric, of
the first derivative along the spin chain.

Summing up over the spin chain sites $k=1,\ldots J$ and passing to the continuous limit,
one finally gets the Hamiltonian of the sigma model~:
\be
\begin{array}{ll}
\bra{\bf n} H \ket{\bf n} &= -\dst{1\over J} \int \dd\sigma\, \dst\,
 g^{AB}\, \partial_\sigma n_A \, \partial_\sigma n_B \\[5mm]
 &=\dst\frac1J \int \dd\sigma\,\bigg(\bar{D}_\sigma\bar{\xi}D_\sigma\xi+\bar{D}_\sigma\bar{\theta}D_\sigma\theta
 +{\bf e}^2\left(1+2\,\bar{\xi}\xi\,\bar{\theta}\theta\right)
 -({\bf \bar{e}_A}{\bf e_A}-{\bf \bar{e}_B}{\bf e_B})(\bar{\xi}\xi+\bar{\theta}\theta)\\[5mm]
 & \hfill+2  \, {\bf \bar{e}_A} \, {\bf \bar{e}_B}\,\theta\,\xi
 +2 \,{\bf e_A} \,{\bf e_B}\, \bar{\xi } \, \bar{\theta }\bigg).
 \end{array}\label{h2}
\ee
In this last expression, the covariant derivative is given by (\ref{WSZ0}) and
\begin{equation}
\begin{array}{l}
{\bf e_A}=\e^{\ii \,\phi}\left(\partial_\sigma\rho
+\frac{\ii}{2}\, \sinh2\rho\ \partial_\sigma\phi\right)~~~~,\quad
{\bf e_B}=\e^{\ii \,\varphi}\left(\partial_\sigma\psi
+\frac{\ii}{2}\, \sin2\psi\ \partial_\sigma\varphi\right)~~,\\[5mm]
{\bf e}={\bf e_A}{\bf \bar{e}_A}+{\bf e_B}{\bf \bar{e}_B}=
(\partial_\sigma \rho )^{2} +\ft14\sinh^2\!2\rho\ (\partial_\sigma\phi )^{2}
+(\partial_\sigma \psi )^{2} +\ft14\sin^2\!2\psi\ (\partial_\sigma\varphi )^{2}~~.
\end{array}
\end{equation}

It is now possible to get the full sigma model action by plugging (\ref{ndn}) and
(\ref{h2}) into (\ref{action}). Its decomposition into bosonic and fermionic
parts writes~:

\begin{equation}
S_B = -J\,\int \dd\sigma\,\dd t \left(-\sinh^2\!{\rho}\ \partial_t \phi
-\sin^2{\psi}\ \partial_t \varphi +{\hat{\lambda}\over J^2} \,  {\bf e}^2\right)~~~~,
\label{finalspinSB}
\end{equation}

\begin{equation}
\begin{array}{l}
\dst S_F = -J\,\int \dd\sigma\dd t \bigg[\frac{\ii}{2}\left(\bar{\xi}\,D_t\xi+\xi\,\bar{D}_t\bar{\xi}
+\bar{\theta}\,D_t\theta+\theta\,\bar{D}_t\bar{\theta}\right)
+{\hat{\lambda}\over J^2} \,\bigg(\bar{D}_\sigma\bar{\xi}D_\sigma\xi
+\bar{D}_\sigma\bar{\theta}D_\sigma\theta\\[5mm]
\hfill\dst +({\bf \bar{e}_B}{\bf e_B}-{\bf \bar{e}_A}{\bf e_A})(\bar{\xi}\xi+\bar{\theta}\theta)
+2  \, {\bf \bar{e}_A} \, {\bf \bar{e}_B}\,\theta\,\xi
 +2 \,{\bf e_A} \,{\bf e_B}\, \bar{\xi } \, \bar{\theta }
 +2\,{\bf e}^2 \bar{\xi}\xi\,\bar{\theta}\theta\bigg)\bigg]~.
\end{array}\label{finalspinSF}
\end{equation}

The bosonic action $S_B$ is exactly the sum of the $SU(2)$ and
$SL(2)$ actions obtained in \cite{Kruczenski:2003gt} and
\cite{Stefanski:2004cw,Bellucci:2004qr}. This structure may appear
as a direct consequence from the fact that the bosonic part of
$su(1,1|2)$ is the direct product $sl(2)\times su(2)$. However,
one should remark that the Hamiltonian projection on bosonic
states (\ref{bosHam}) is not the sum of the Hamiltonians
restricted to these two subsectors. As it is proved in section
\ref{sstring}, $S_B$ corresponds to the bosonic action of a string
spinning fast in $S_{\phi_1}\times S_{\varphi_1}\times
S_{\varphi_2}$, with $S_{\phi_1}$ in AdS$_5$ and
$S_{\varphi_1}\times S_{\varphi_2}$ in $S^5$.

The action $S_F$ appears to be more complicated, as it is through
the fermions that $SU(2)$ and $SL(2)$ sectors interact. As one
could have expected, it is not quadratic in fermions, although the
quartic term in Grassmann variables is just proportional to the
bosonic Hamiltonian ${\bf e}^2$. Because of the mixed coefficients
in front of the Grassmann variables, it seems that getting the
corresponding superstring description will be but a hard task. For
example, in the $SU(2|3)$ sector were one deals with three complex
scalars and two complex fermions, the Grassmann variables appear
in the sigma model with the same factor ${\bf e}^2$
\cite{Hernandez:2004kr}.

\section{Bosonic string action}
\label{sstring}

As one can expect from the bosonic action (\ref{finalspinSB}), the
``fast spinning'' limit to the dual bosonic string is a mix of the
$SU(2)$ and $SL(2)$ found limits
\cite{Kruczenski:2003gt,Stefanski:2004cw,Bellucci:2004qr}. What
happens here is that $SU(2)$ and $SL(2)$ parts will share a fast
spinning circle in $S^5$.

The bosonic part of Polyakov action describing a string moving on
 $AdS_5\times  S^5$ can be written as
 \begin{eqnarray}
 S={R^2\over 4 \pi \alpha^\prime} \int g_{MN} (\partial_\tau X^M
\partial_\tau
 X^N-\partial_\sigma X^M \partial_\sigma X^N)\label{straction}
 \end{eqnarray}
 with
 $$
\dd s^2 = g_{MN} dX^M dX^N=\dd s^2_{\rm{AdS}_5}+\dd s^2_{S^5}
 $$
 and
\begin{eqnarray}
\dd s^2_{\rm{AdS}_5}&=&\dd \rho^2-\cosh^2 \rho\, \dd t^2+\sinh^2 \rho\,
 (\dd \theta^2+\cos^2  \theta \,d\phi_1^2+\sin^2 \theta\, \dd \phi_2^2)~,\nn\\
 \dd s^2_{S^5}&=&\dd \gamma^2+\cos^2 \gamma\, \dd \varphi_3^2+\sin^2
 \gamma\,(\dd  \psi^2+\cos^2  \psi\, \dd \varphi_1^2+\sin^2 \psi
 \,\dd \varphi_2^2)~.\label{metric}
 \end{eqnarray}
Imposing $\gamma=\ft\pi2$, $\theta=0$ and making the change of variables
\begin{equation}
 \phi_1=\phi+t~, \qquad \varphi_1=\hat\varphi+t~,\qquad \varphi_2=\varphi+\hat\varphi+t~,
\end{equation}
the metric (\ref{metric}) rewrites as
\begin{eqnarray}
\dd s^2&=&2\,\dd t\,\left(\dd \hat\varphi+\sinh^{2}\!\!\rho\ \dd \phi
+\sin^{2}\!\!\psi\ \dd \varphi\right )+\dd \hat\varphi^{2}
+2\sin^{2}\!\!\psi\ \dd \varphi\,\dd\hat\varphi\nn\\
&&+\dd \rho ^{2}+\sinh^{2}\!\!\rho \ d\phi^{2}+\dd\psi^{2}+\sin^{2}\!\!\psi\ \dd \varphi^{2}~.
\label{metricnew}
\end{eqnarray}
As usual, we make the light-cone gauge choice $t=\kappa\, \tau$
and take the limit $\kappa \rightarrow +\infty$, keeping
$\kappa\,\partial_{\tau} X^M$ fixed for the other coordinates.

~\newline
The action (\ref{straction}) should also satisfy the Virasoro constraints.
To leading order in $\kappa$, the first of them reads
\begin{equation}
g_{MN} \,\partial_\tau X^M \partial_\sigma X^N=
\kappa\, \left(
\sin^{2} \!\!\psi\ \partial_\sigma \varphi +\sinh^{2}\!\!\rho\ \partial_\sigma \phi
+\partial_\sigma\hat\varphi
\right)=0\, ,
  \label{vir1}\end{equation}
and can be used to
solve for $\partial_\sigma\hat\varphi $.

Evaluating the action (\ref{straction}) with the metric (\ref{metricnew}) and
using (\ref{vir1}), one gets  to leading order in $\kappa$
\begin{eqnarray}
   S &=&-{R^2 \,\kappa\over 2 \pi
   \alpha^\prime} \int \dd\sigma \dd t \bigg(-\sinh^2\!{\rho}\ \partial_t \phi
             -\sin^2{\psi}\ \partial_t \varphi -\partial_t \hat\varphi\nn\\
   &&\quad\quad\quad\quad\quad\quad\quad\quad+\frac1{2\,\kappa^2}
   \,\big[(\partial_\sigma\rho)^2+ \ft14 \sinh^2\!\!\rho\,(\partial_\sigma\phi)^2
           +(\partial_\sigma\psi)^2+ \ft14 \sin^2\!\!\psi\,(\partial_\sigma\varphi)^2
   \big]\bigg) \, .\nn
\end{eqnarray}
Identifying
\begin{equation}
J={R^2 \kappa \over 2 \pi \alpha^\prime}\qquad\rm{and}\quad
\tilde{\lambda}={R^4\over 8\, \pi^2 \alpha^{\prime 2}}~,
\end{equation}
the string action gives back the bosonic spin chain sigma model (\ref{finalspinSB}),
up to the full time derivative $\partial_t \hat\varphi$.

In the original variables, the limit $\kappa \rightarrow +\infty$
corresponds to $\partial_\tau \phi_1\approx\partial_\tau \varphi_1
\approx \partial_\tau \varphi_2\approx k$, so the string spins
fast on $S^1_{\varphi_1}\times S^1_{\varphi_2} \in S^5$ and
$S^1_{\phi_1}\in\rm{AdS}_5$. The most simple non trivial solutions
for the classical equations of motion is then given by a
multi-spinning string folded or circular by respect to the $\psi$
coordinate \cite{Kruczenski:2003gt} and folded by respect to the
$\rho$ coordinate \cite{Bellucci:2004qr}.

\section{Conclusion}
\label{scon}

The sigma model arising from $SU(1,1|2)$  spin chains was derived. Doing so,
a truncated Hamiltonian on $SU(2)\times SL(2)$ spin chains appeared. This truncated
Hamiltonian leads to a one to one correspondence between long bosonic spin chains states
in the $SU(1,1|2)$ planar subsector of ${\cal N}=4$ SYM gauge theory
at one loop, and bosonic strings spinning fast on two circles in $S^5$ and one circle in
AdS$_5$. The resulting action is the sum of the sigma models arising from $SU(2)$ spin chains
and $SL(2)$ ones. The average of this Hamiltonian between two neighboring bosonic coherent states
 is not anymore
expressible in terms of the full ``coherent vector'' square $(\vec{n}_{k+1}-\vec{n}_{k})^2$ but
rather as a mixing between pure $SU(2)$ and $SL(2)$ terms~:
$$\bra{\bf n} H^{\rm{bosonic}} \ket{\bf n} =\sum_{k=1}^L\, \left(1-\frac {(\vec{n}_{k+1}-\vec{n}_{k})_{su2}^2}
{(\vec{n}_{k+1}-\vec{n}_{k})_{sl2}^2}\right)\,\log(1-(\vec{n}_{k+1}-\vec{n}_{k})_{sl2}^2)~~.$$
When one takes into account the fermionic part, the correspondence to super-strings seams much
more involved. Such difficulties should increase in enlarging to bigger sectors of the full theory,
and a way out could be to build the sigma models not in terms of precise coordinates, but rather
in terms of more general expressions with constraints given by the coset structure as proposed
in \cite{Stefanski:2005tr}. Another possibility could be to reason in terms of Cartan forms $L^a$, as
the string action on AdS$_5\times S^5$ in terms of these is known \cite{Metsaev:1998it}. Indeed,
as soon as the Hamiltonian in the continuous limit is proportional to $(\pd_\sigma\vec{n})^2$,
as it is the case here,
it is possible to express the spin chain sigma model in a $G$-invariant form as \cite{Bellucci:2005vq}
$$S= -J\,\int \dd ^2\sigma\left[\ii\,L^A_t\, n_A-
 {\hat{\lambda}\over J^2}\, (L_{\sigma}^B\, {f_{BA}}^C\, n_C )^2\right],$$
where ${f_{BA}}^C$ are the structure constants of the considered group $G$.

The two-loop Hamiltonian of $SU(1,1|2)$ sector was given recently in
\cite{Zwiebel:2005er} in terms of rising/lowering length operators. At higher loops,
the Hamiltonian starts to change the length of the spin chain. However, such interactions may
be absent in the continuous limit, as argued in \cite{Minahan:2004ds}. Then it would be possible
to compute the sigma model up to two loops and see how it would match,
at least for the bosonic part, with the fast spinning string.

Another issue is to ask oneself the integrability of the truncated, bosonic, Hamiltonian
(\ref{bosHam}). Although we checked that the simple Ansatz
$Q=\sum_{k=1}^L[H_{k+2,k+1}^{\rm bosonic},H_{k+1,k}^{\rm bosonic}]$ for the next higher charge
does not commute with $H^{\rm bosonic}$, integrability may be retained in some involved way.
It would then provide us an example of an integrable $SU(2)\times SL(2)$ spin chain.

\subsection*{Acknowledgements}
We thank Francisco Morales for very useful and animated discussions.
This research was partially supported by the European
Community's Marie Curie Research Training Network under contract
MRTN-CT-2004-005104 Forces Universe.
\appendix
\section*{Appendix}
\renewcommand{\theequation}{A.0}

In this appendix we collect the commutation relations and details
on the ``singleton" representations of  the superalgebra
$\alG=\alSU(1,1|2)$. A singleton corresponds
to a subsector of the ${\cal N}=4$ SYM multiplet that closes under
$\alG$. Here we adopt the oscillator description (see \cite{Beisert:2003jj} for
details). In this formalism, elementary SYM fields (the singleton
of $\alPSU(2,2|4)$) are represented by acting on a Fock vacuum
$|0\rangle$ with bosonic $(a_\alpha, b_{\dot{\alpha}})$ and
fermionic oscillators $c_A$,
($\alpha,\dot{\alpha}=1,2, A=1,\ldots 4$). Physical states
 satisfy the condition
\begin{equation}
C=n_a-n_b+n_c =2 \label{c0}
\end{equation}
with $n_a,n_b,n_c$ denoting the number of oscillators of a given type.

 The closed subalgebras of $\alSU(2,2|4)$ are defined by
restricting the range of $\alpha,\dot{\alpha},A$.

\renewcommand{\theequation}{\Alph{section}.\arabic{equation}}
\section{$\alSU(1,1|2)$ algebra}
\label{aa}

 The algebra $su(1,1|2)$ is built in terms of bilinears  of two
bosonic ($a$, $b$) and two fermionic ($c_2$, $c_3$) oscillators.
It consists of an $su(2)$ charge $R_0$, an $sl(2)$ charge $J_0$,
Lorentz translation and boost $P$ and $K$, $su(2)$ R-symmetry
rotation operators $R_{23}$ and $R_{32}$, and eight fermionic
supertranslations and superboosts $Q_i$, $\dot{Q}_i$,
$S_i$ and $\dot{S}_i$. Here, $i=2,3$, in order to follow the notations of \cite{Beisert:2003jj}. We choose as physical vacuum the state
$|\phi_0\rangle=c^\dagger_1c^\dagger_2|0\rangle$.

States in the singleton representation are given by
\begin{equation}
\label{stasu112}
\begin{array}{rll}
 |\phi_m\rangle =&\dst \frac{1}{m!}(a\dg b\dg)^m |\phi_0\rangle
&\dst \Leftrightarrow ~~{1\over m!}\, {\cal D}^m \phi_0 \\[2mm]
 |Z_m\rangle =&\dst \frac{1}{m!}(a\dg b\dg)^m c_3\dg c_2 |\phi_0\rangle
&\dst \Leftrightarrow ~~{1\over m!}\, {\cal D}^m Z_0 \\[2mm]
 |\lambda_m\rangle =&\dst \frac{1}{m!}(a\dg b\dg)^m\, b\dg c_3\dg
|\phi_0\rangle &\dst \Leftrightarrow ~~{1\over m!}\,{\cal D}^m \lambda_0\\[2mm]
 |\mu_m\rangle =&\dst \frac{1}{m!}(a\dg b\dg)^m\, a\dg c_2
|\phi_0\rangle &\dst \Leftrightarrow ~~{1\over m!}\,{\cal D}^m \mu_0
\end{array}
\end{equation}
 and correspond to two scalar fields $\phi_0$, $Z_0$ and two fermions $\lambda_0$, $\mu_0$, together
with their $m$-derivatives along a fixed direction. In order to get rid of square roots in
expressions, the states are normalized according to~:
$$\langle\phi_m|\phi_m\rangle=\langle Z_m|Z_m\rangle=1\,,~~~~
\langle\lambda_m|\lambda_m\rangle=\langle\mu_m|\mu_m\rangle=m+1~~.$$
 The algebra in this case is
non-compact and the representations are infinite-dimensional.

The generators in terms of oscillators are given by
 \bea \begin{array}{l}
R_{23}=c_2\dg c_3\\[2mm]
R_{32}=c_3\dg c_2\\[2mm]
P=a\dg b\dg \\[2mm]
K=a\,b
\end{array}
\quad\quad
\begin{array}{l}
Q_i=a\dg c_i\\[2mm]
\dot{Q}_i=b\dg c_i\dg\\[2mm]
S_i=a\, c_i\dg\\[2mm]
\dot{S}_i=b\, c_i
\end{array}
\quad\quad
\begin{array}{l}
\\[2mm]
J_0=\ft12(1+a\dg a+b\dg b)\\[2mm]
R_0=\ft12(c_2\dg c_2 -c_3\dg c_3)\\[2mm]

\end{array} \nn
\eea The $sl(2)$ and $su(2)$ charges $J_0$ and $R_0$ give the Cartan of the group. Non
vanishing commutation relations are given in the Tables \ref{table:ferfer},
\ref{table:bosfer}, \ref{table:bosbos}. The action of the charges upon states is given
in Table \ref{table:actstates}.

\begin{table}[p]
$$\begin{array}{|c||ccccccc|}\hline
\{\downarrow,\rightarrow\}&S_2&Q_3&S_3&\dot{Q}_2&\dot{S}_2&\dot{Q}_3&\dot{S}_3\\ \hline\hline
Q_2     & J_0+R_0   &       &  R_{32} & P &       &   &   \\
S_2     &           &R_{23} &         &   & K     &   &      \\
Q_3     &      &       & J_0-R_0 &   &       & P &      \\
S_3     &           &  &         &   &       &   &  K   \\
\dot{Q}_2&          &       &         &   &J_0-R_0&   & -R_{23} \\
\dot{S}_2&     &       &         &&     &-R_{32}&      \\
\dot{Q}_3&          &  &         &   &  &       &  J_0+R_0   \\ \hline
\end{array}$$
\caption{Fermionic anticommutators. Here and below,
redundant subdiagonal terms are omitted.
\label{table:ferfer}}
\end{table}

\newcommand{\phm}{\phantom{-}}
\begin{table}[p]
$$\begin{array}{|c||cccccccc|}\hline
{[}\downarrow,\rightarrow]&Q_2&S_2&Q_3&S_3&\dot{Q}_2&\dot{S}_2&\dot{Q}_3&\dot{S}_3\\ \hline\hline
P      &              & -\dot{Q}_2 &             & -\dot{Q}_3 &              & -Q_2          &                  & -Q_3 \\
K      & \phm\dot{S}_2&            &\phm\dot{S}_3&            &  S_2         &               & \phm S_3         &  \\
R_{23} & -Q_3         &            &             & \phm S_2   &              &-\dot{S}_3     & \phm\dot{Q}_2    &     \\
R_{32} &              & \phm S_3   &- Q_2        &            &\dot{Q}_3     &               &                  &-\dot{S}_2   \\
J_0    & \phm\ft12 Q_2&-\ft12 S_2  &\phm\ft12 Q_3&-\ft12 S_3  &\ft12\dot{Q}_2&-\ft12\dot{S}_2&\phm\ft12\dot{Q}_3&-\ft12\dot{S}_3 \\
R_0    & -\ft12 Q_2   &\phm\ft12S_2&\phm\ft12 Q_3&-\ft12 S_3  &\ft12\dot{Q}_2&-\ft12\dot{S}_2&-\ft12\dot{Q}_3   &\phm\ft12\dot{S}_3\\\hline
\end{array}$$
\caption{Fermionic/Bosonic commutators\label{table:bosfer}}
\end{table}

\begin{table}[p]
$$\begin{array}{|c||ccccc|}\hline
{[}\downarrow,\rightarrow]&K&R_{23}&R_{32}&J_0&R_0\\ \hline\hline
P      & -2 J_0 &   &  & -P &  \\
K      &        &   &  & \phm K &  \\
R_{23} &        &   & 2R_0 &  & -R_{23} \\
R_{32} &        &  &  &  &\phm R_{32} \\\hline
\end{array}$$
\caption{Bosonic commutators\label{table:bosbos}}
\end{table}

\newcommand{\mcc}{\multicolumn{2}{c|}}
\newcommand{\mch}{&\hspace*{-.3cm}}
\begin{table}[p]
$$\begin{array}{|c||rl|rl|rl|rl|}\hline
\downarrow \ket{\rightarrow}&\mcc{\ket{\phi_m}}&\mcc{\ket{Z_m}}&\mcc{\ket{\lambda_m}}&\mcc{\ket{\mu_m}}               \\\hline\hline[2mm]
Q_2      &\mch\ket{\mu_m}           &\mch0                   &-(m+1)\mch\ket{Z_{m+1}}     &\mch0                     \\[2mm]
S_2      &\mch0                      &-\mch\ket{\lambda_{m-1}}&\mch0                       &(m+1)\mch\ket{\phi_{m}}  \\[2mm]
Q_3      &\mch0                      &\mch\ket{\mu_{m}}      &(m+1)\mch\ket{\phi_{m+1}}   &\mch0                     \\[2mm]
S_3      &\mch\ket{\lambda_{m-1}}    &\mch0                   &\mch0                       &(m+1)\mch\ket{Z_{m}}     \\[2mm]
\dot{Q}_2&\mch0                      &-\mch\ket{\lambda_{m}}  &\mch0                       &(m+1)\mch\ket{\phi_{m+1}}\\[2mm]
\dot{S}_2&\mch\ket{\mu_{m-1}}       &\mch0                   &-(m+1)\mch\ket{Z_{m}}       &\mch0                     \\[2mm]
\dot{Q}_3&\mch\ket{\lambda_{m}}      &\mch0                   &\mch0                       &(m+1)\mch\ket{Z_{m+1}}   \\[2mm]
\dot{S}_3&\mch0                      &\mch\ket{\mu_{m-1}}    &(m+1)\mch\ket{\phi_{m}}     &\mch0                     \\[2mm]
P        &(m+1)\mch\ket{\phi_{m+1}}  &(m+1)\mch\ket{Z_{m+1}}  &(m+1)\mch\ket{\lambda_{m+1}}&(m+1)\mch\ket{\mu_{m+1}}  \\[2mm]
K        &m\mch\ket{\phi_{m-1}}      &m\mch\ket{Z_{m-1}}      &(m+1)\mch\ket{\lambda_{m-1}}&(m+1)\mch\ket{\mu_{m-1}}  \\[2mm]
R_{23}   &\mch0                      &\mch\ket{\phi_{m}}      &\mch0                       &\mch0                     \\[2mm]
R_{32}   &\mch\ket{Z_{m}}            &\mch0                   &\mch0                       &\mch0                     \\[2mm]
J_0      &(m+\ft12)\mch\ket{\phi_{m}}&(m+\ft12)\mch\ket{Z_{m}}&(m+1)\mch\ket{\lambda_{m}}  &(m+1)\mch\ket{\mu_{m}}    \\[2mm]
R_0      &\ft12\mch\ket{\phi_{m}}    &-\ft12\mch\ket{Z_{m}} &\mch0                       &\mch0                     \\\hline
\end{array}$$
\caption{Action of $SU(1,1|2)$ charges\label{table:actstates}}
\end{table}

A single trace SYM operator of length $J$ is given by the tensor products of $J$ singletons~: we
take J copies of the oscillators $a,b,c$ and impose the condition
(\ref{c0}) at each site.
The symmetry algebra is then taken to be the diagonal SU$(1,1|2)$ algebra
\be
 \tilde T_A=\sum_{k=1}^J\, (T_A)_k
\ee
 with $(T_A)_k$ acting on the $k^{th}$ site.

The Killing metric of $SU(1,1|2)$ vanishes identically. However,
it is still possible to define a metric $g_{AB}$ through the
Casimir of the algebra, which is given here by~:
\begin{equation}\label{casimir}
  \hat{C}_2= g^{AB}\,T_A T_B= J_0^2- R_0^2-\frac12\{P,K\}-\frac12\{R_{23},R_{32}\}
  -\frac12[Q_i,S_i]-\frac12[\dot{Q}_i,\dot{S}_i]~.
\end{equation}
$\hat{C}_2$ defines the spin $j$ by
$$\hat{C}_2=j~(j+1)~ \I~.$$
In the singleton representation, the spin vanishes : $j=0$ for all the 1-site states (\ref{stasu112}).
Spin $j$ representations arise then in the tensor product
of two singletons and the corresponding highest weight states can be written as follows~:
\be
 |j\,\rangle_{1,2}=\sum_{i=0}^j \,(-1)^i\,\left(\!\!
\begin{array}{c}
  j \\
  i \\
\end{array}%
\!\!\right)\, |\phi_{j-i}\rangle_{1} \otimes |\phi_{i}\rangle_{2}~.\label{spinj}
\ee



\providecommand{\href}[2]{#2}\begingroup\raggedright\endgroup

\end{document}